# Switchable enhanced spin photocurrent in Rashba and cubic Dresselhaus ferroelectric semiconductors

*Ruixiang Fei,* [†,*] *Shuaiqin Yu,* [‡] *Yan Lu,* [†, §] *Linghan Zhu,* [†] *Li Yang* [†, ǁ,*]

[†] Department of Physics, Washington University in St Louis, St Louis, Missouri 63130, United States

[‡] College of Ocean Science and Engineering, Shanghai Maritime University, Shanghai 201306, PR China

[§] Department of Physics, Nanchang University, Nanchang, 330031, PR China

[ǁ] Institute of Materials Science and Engineering, Washington University in St. Louis, St. Louis, Missouri 63130, United States

ABSTRACT

Generating and controlling spin current (SC) are of central interest in spin physics and applications. To date, the spin-orbit interaction (SOI) is an established pathway to generate SC through the spin-charge current conversion. We predict an efficient spin-light conversion via the Rashba and higher-order cubic Dresselhaus SOIs in ferroelectrics. Different from the known Edelstein effect, where SC is created by the nonequilibrium spin density, our predicted spin-polarized current is from direct interactions between light and unique spin textures generated by SOI in ferroelectrics. Using first-principles simulations, we demonstrate these concepts by calculating the DC spin photocurrent in a prototypical Rashba ferroelectric, α-GeTe. The photo-induced SC is about two orders of magnitude larger than the charge photocurrent. More importantly, we can conveniently switch the direction of SC by an applied electric field via inverting the spin textures. These predictions give hope to generating and controlling light-driven SC via nonvolatile electric-field.

**Introduction.**

The nonvolatile electric control of spin-charge conversion in materials is highly desired to explore spin physics and applications because it holds promise to combine information storage and computing functionalities. The spin-orbit interaction (SOI) has been identified as a fundamental pathway for this purpose because it can induce spin-polarized currents from charge currents and versa [1–3]. The two phenomena at the core of spintronics are the spin-Hall effect and the inverse spin Hall effect [4,5]. Alternatively, the Rashba SOI leads to a spin locking of electrons according to their momenta, enabling efficient spin-to-charge current interconversions through Edelstein [6,7] and inverse Edelstein effects [8–10]. To date, the nonvolatile electric control of spin in giant Rashba SOI semiconductors, e.g. GeTe [11,12], BiTeI [13,14], transition-metal perovskites [15], and organometal halide perovskites [16–18], as well as the SrTiO3 Rashba electron-gas system [19,20], has spurred tremendous research interest. In these materials, the spin degree of freedom is inherently coupled to the electric polarization, whereby the polar axis intrinsically breaks the inversion symmetry. As a result, the Rasha spin-splitting of the bulk electronic states can be then switched with the ferroelectric orientation and subsequently controlled by an electric field [11,12,19].

Beyond the static electric field, the dynamic electromagnetic field, i.e., light, can be another promising approach to tune the spin current of Rashba SOI systems, because it can also generate a DC spin-current in non-centrosymmetric materials with SOI, such as CdSe and GaAs quantum well structures [21–23]. More recently, the second-order of AC electric field was proposed to generate the charge current due to the non-zero Berry curvature(BC) dipole, namely the nonlinear Hall effect [24]. Using the concept of BC dipole and semiclassic method, the spin current and charge current were predicted in mono-layer SnTe under circularly polarized light [25]. However, in those nonmagnetic materials, the spin current is reported to be one to two orders of magnitude smaller than the charge current excited by the circularly polarized light [26,27]. As a result, it is challenging for light to generate the highly desired pure spin current (PSC), in which electrons with different spins travel in opposite directions while there is no net motion of charge.

In this work, we propose that linearly polarized light can generate enhanced PSC in Rashba and cubic Dresselhaus SOI systems. Different from the Edelstein effect, where the induced current is created by the nonequilibrium spin density, the predicted in-plane spin-polarized current is from the opposite group velocities of spins associated with Rashba SOI. We demonstrate these concepts by directly calculating the DC spin current in a known Rashba ferroelectric material, bulk α-GeTe. Furthermore, we find the off-plane spin photocurrents due to the hidden cubic Dresselhaus SOI of α-GeTe. Moreover, the direction of PSC can be conveniently tuned by the ferroelectric polarization of the materials. This prediction gives hope for generating and tuning PSC by light in an intrinsic non-centrosymmetric crystal.

**Results and Discussion.**

Previous works have shown that a PSC can be generated by a single-frequency linearly polarized light in CdSe and GaAs quantum well because of the opposite transport behavior of injected non-equilibrium spin-polarized carriers[26,28,29]. From the point of view of the quantum perturbation theory for nonlinear light-matter interactions, the photo-induced spin currents under linearly and circularly polarized light are described as the shift and injection currents, respectively [30–32]. In the following, we present the SOI Hamiltonian and corresponding photocurrent calculated by the quantum perturbation theory. It has to be addressed that the main results are based on the symmetry and associate SOI spin textures, which are not sensitive to the particular calculating approach of photocurrent.

***Linear polarization induced spin photocurrent.*** In materials lacking inversion symmetry, the interplay between the electric field and SOI lifts the spin degeneracy of electronic bands. For a material exhibiting a polar axis oriented along the z-direction, we investigate the photo-driven spin current by examining the spin splitting by the SOI Hamiltonian[16,33]

$$H_R\left(\vec{k}\right) = \frac{\hbar^2 \vec{k}^2}{2m} I_{2x2} + \hbar\lambda_R \left(\hat{z} \times \vec{k}\right) \cdot \vec{\sigma} + O(k_+^3, k_-^3). \tag{1}$$

Where $k_\pm = k_x \pm ik_y$ and $\vec{\sigma}$ are the Pauli matrices; The parameters $m$ and $\lambda_R$ represent the band mass and Rashba SOI strength, respectively. The higher-order SOI terms $O(k_+^3, k_-^3)$ are enabled in particular symmetries of the structure without inversion symmetry. Here, we firstly focus on the scenario that the Rashba term is the leading order, and the spin up and down energy spectra are

$$E\left(\vec{k}\right)_\pm = \frac{\hbar^2 k_z^2}{2m} + \frac{\hbar^2 (k_x^2 + k_y^2)}{2m} \pm \hbar\lambda_R \sqrt{k_x^2 + k_y^2}. \tag{2}$$

Figure 1a shows the conduction bands and valence bands of the Rashba Hamiltonian. Near the bandgap, the Rashba band edges are formed, and spin degeneracies of the conduction and valence bands are lifted, giving rise to the "outer" and "inner" bands with opposite spin textures. The spin rotation directions are characterized as "clockwise" (blue bands) and "counterclockwise" (red bands) accordingly. There are two sets of spin textures for the conduction and valence bands: 1) both "outer" conduction and valence bands exhibit the "clockwise" spin textures, which is depicted in Figure 1a, or 2) the "outer" conduction and "inner" valence bands show the "clockwise" spin textures. It is worth mentioning that the photo-driven spin current, in general, is not affected by this relative spin-texture relation between the conduction bands and valence bands (see Figure S1 of SI). In the following discussion, we use the former case which matches our calculated spin textures of real material, bulk $\alpha$-GeTe.

Next, we use quantum perturbation theory to calculate the second-order photo-induced current. Generally, two nonlinear photocurrent mechanisms, i.e. shift current and injection current, contribute to the bulk photovoltaic effect[34–37]. Under linearly polarized light, although the injection current mechanism contributes to a zero charge current, it has been recently predicted that it can generate non-zero or even enhanced PSC[32]. Using the free-carrier relaxation time approximation, the DC photoconductivity $\eta^{\alpha}_{abc}$ of injection current ($j^{\alpha}_c = \eta^{\alpha}_{abc} E_a(\omega) E_b(-\omega)$) is

$$\eta^{\alpha}_{abc} = \frac{-\pi e^3}{\hbar^2 \omega^2} \sum_{mn} \int d^3k \, \alpha^{ab}_{mn}(k) \big(\langle m|\{\sigma_\alpha, v_c\}|m\rangle \tau(E_{k_m}) -$$

$$\langle n|\{\sigma_\alpha, v_c\}|n\rangle \tau(E_{k_n})\big) \delta(\omega - \omega_{mn}). \quad (3)$$

where $\alpha^{ab}_{mn}(k) = \frac{1}{2}\big(v^a_{mn} v^b_{nm}(k) + v^b_{mn}(k) v^a_{nm}(k)\big)$ is the real part of the optical oscillator strength. $v^a_{mn}(k)$ and $v^b_{nm}(k)$ are the $a$-direction and $b$-direction interband velocity matrices between the conduction band $m$ and the valence band $n$ for the same wave vector $k$, respectively. $\langle m|\{\sigma_\alpha, v_c\}|m\rangle$ is the $c$-direction velocity of the $m$-band electron with the $\alpha$-direction spin, and $\{\sigma_\alpha, v_c\} \equiv \frac{1}{2}(v_c \sigma_\alpha + \sigma_\alpha v_c)$. Thus, the net spin current has both contributions of holes and electrons. $\tau(E_k)$ is the minimum value of the spin-relaxation time and free-carrier relaxation time, which dependens on the band energy ($E_k$).

For simplicity, we use the "outer" conduction or valence band contour in Figure 1b to demonstrate the mechanism. When pumped by the frequency $\omega$ photon, the spin-up electrons travel towards the left ($-k_x$) while the spin-down electrons travel towards the right ($k_x$), forming a spin-polarized current indicated by Figure 1b. Although the "inner" band transition may contribute to an opposite-direction spin current, it cannot cancel the "outer" band transition contribution because of their different group velocities and oscillator strengths induced by Rashba

SOI. Thus, the spin-up and spin-down carriers with opposite group velocities induce an overall non-zero spin-polarized photocurrent. We have to address that this mechanism is fundamentally different from the known Edelstein effect, in which the spin current is generated through the shift of the Fermi-level contours from equilibrium and the subsequent non-equilibrium electron spin density.

***Rashba SOI and cubic Dresselhaus SOI in bulk α-GeTe.*** In the following, we take bulk $α$-GeTe to demonstrate and quantify this physics picture. Bulk α-GeTe is a well-known non-centrosymmetric material with giant Rashba SOI as reported in theory and experiment[11,12]. The α-GeTe, below a high critical temperature $T_c \approx 720\ K$, stabilizes in a ferroelectric structure with the space group R3m (No. 160), in which Ge and Te ions are displaced from the ideal rocksalt sites along the (111) direction, as shown in Figure 2a. Figure 2b present its first Brillouizon zone and high-symmetry points. The band dispersion along the high-symmetric lines is displayed in Figure 2c. Since the symmetry breaking field is along the z-direction, according to Eq. (2), there is no spin splitting along the $\Gamma T$ line while a giant spin splitting is observed on the $k_z = \frac{\pi}{2}$ plane, namely the hexagonal $TMU$ plane. Figures 2d and 2e show the energy contour of the valence band at -0.1 eV and that of the conduction band at 0.84 eV, respectively. The spin textures of these contours suggest that the spin-splitting is dominated by the Rashba SOI. Both the spin textures of the "outer" valence band and conduction band are "clockwise", which corresponds to the case depicted in Figure 1.

The Rashba type SOI is viewed as the dominant contribution to the spin splitting[11], and the higher-order SOI is overlooked in this kind of material. However, high-order SOI such as the cubic Dresselhaus-type SOI can also exist and cause exotic phenomena. We find that the observed anisotropic energy contour can be explained by higher-order SOI terms in the Hamiltonian. Similar to the topological insulator $Bi_2Se_3$ [38], the symmetry of the (111) surface of bulk α-GeTe is reduced to $C_{3v}$. Thus, the cubic Dresselhaus SOI is allowed and the $H(\vec{k})$ in Eq. 1 can be extended to the third order of the wavevector $\vec{k}$:

$$H(\vec{k}) = \frac{\hbar^2 \vec{k}^2}{2m} I_{2x2} + \hbar\lambda_R(\hat{z} \times \vec{k}) \cdot \vec{\sigma} + \lambda_{cd}(k_+^3 + k_-^3)\sigma_z \qquad (4)$$

where the last term is cubic Dresselhaus SOI and the $\lambda_{cd}$ is the cubic Dresselhaus SOI strength. Such a term is crucial for generating the z-component photocurrent since the Rashba SOI does not induce the *z*-component spin splitting. To view the role of Dresselhaus SOI, we show the energy contour of the valence and conduction bands as well as the spin $S_z$ in Figures 2d and 2e. The snowflake-like energy contour of the 'outer' band is attributed to the cubic Dresselhaus SOI which is only invariant under the three-fold rotation operation. Particularly, since $k_+^3$ and $k_-^3$ are small for the 'inner' band, its energy contour is dominated by the Rashba SOI, exhibiting a circular surface. As a result, the *z*-component pseudo-spin of the 'inner' band contour is nearly zero while that of the 'outer' band cannot be neglected.

***Enhanced photo-driven PSC in bulk α-GeTe.*** The bulk α-GeTe is expected to be a good candidate for photovoltaic devices because it has a sizable bulk photovoltaic effect with a narrow bandgap (0.6–0.7 eV)[39] and strong ferroelectric polarization ($P_s = 60 - 70 \ \mu C/cm^2$)[40]. In the following, we will show that the injection current using Eq. 3, other than the aforementioned bulk photovoltaic effect, can contribute to enhanced spin current under linearly polarized light. Briefly, the shift current and injection current are determined by non-diagonal and diagonal matrix elements of the current operator in Bloch basis[35,41], respectively. Usually, the magnitude of shift current is one to two orders of magnitude smaller than that of the injection current[42]. Thus, we expect the injection spin current of bulk α-GeTe generated by linearly polarized light to be at the order of $mA/V^2$ within the visible spectrum, compared with the reported charge current photoconductivity (tens of $\mu A/V^2$) [40].

We employ first-principles calculations to obtain the PSC spectra of bulk α-GeTe. Figure 3a shows the photo-driven PSC of the z-component pseudo-spin ($S_z$). We calculate the current photoconductivity by Eq. 3, and $\eta_{abc}^{Sz} = \eta_{abc}^{Sz}(Sz > 0) - \eta_{abc}^{Sz}(Sz < 0)$ is defined to be the net spin current. Meanwhile, we confirm that the charge current photoconductivity $\eta_{abc}^{Sz}(Sz > 0) + \eta_{abc}^{Sz}(Sz < 0)$ is zero, which is consistent with the well-known result that the injection current is zero under linearly polarized light for time-reversal symmetric materials [35]. It is worth mentioning that the calculation of the above inject-current spectra needs an electron lifetime $\tau(E_m)$ and hole lifetime $\tau(E_n)$, as indicated by Eq. 3. For example, the hole mobility of α-GeTe was measured to be around 100 $cm^2/V \cdot s$ at room temperature[43], which is comparable with those widely studied transition metal dichalcogenides (TMDs). Given that the carrier lifetime of MoS$_2$ is $\tau = 1$ ps [44], we choose a conservative value, $\tau = 0.1$ ps, which is an order of less. We also assume the electron and hole lifetimes are the same and set them to be a constant for different band energies. As shown in Figure 3a, the $S_z$ component photoconductivity $\eta_{xxY}^{S_z}$ can reach up to 1.4 $mA/V^2$ for incident photons around 0.9 eV.

Impressively, such a giant photo-driven spin current is around two orders of magnitude larger than that of the previously calculated charge current[40], agreeing with the aforementioned analysis and resulting in the predicted PSC. Moreover, the magnitude of these spin currents is of the same order as the charge current of ferroelectrics under circularly polarized light[27,45], which is approximately one to two orders of magnitude larger than the PSC generated in CdSe and GaAs quantum well structures[21–23]. It is worth mentioning that, in addition to SOI, the dispersive *sp*-character conduction and valence bands of bulk GeTe lead to high group velocity carriers and enhance this giant spin current.

However, the *z*-component pseudo-spin should be zero if we only consider the Rashba SOI, as discussed in the previous section. Our further analysis shows that this non-zero *z*-component PSC ($\eta_{abc}^{S_z}$) is attributed to the cubic Dresselhaus SOI. Due to the cubic Dresselhaus SOI as presented in Eq. 4, which is enforced by the three-fold rotation $C_3$ and mirror operation, the nontrivial PSC photoconductivity tensor elements satisfy $\eta_{xxY}^{S_z} = \eta_{xyX}^{S_z} = \eta_{yxX}^{S_z} = -\eta_{yyY}^{S_z}$, and $\eta_{xzY}^{S_z} = \eta_{zxY}^{S_z} = -\eta_{yzX}^{S_z} = -\eta_{zyX}^{S_z}$. Figure 3b presents the contour plot of the calculated $S_z$ component photoconductivity $\eta_{xxY}^{S_z}$ on the (111) plane of the reciprocal space. This plot clearly demonstrates the role of the cubic Dresselhaus SOI in generating photo-driven spin current. The 'inner' circular band contour is dominated by the Rashba SOI, hence have few contributions to the spin current, whereas the cubic Dresselhaus SOI for the 'outer' band contour cannot be neglected and contributes to the sizeable $S_z$ component of spin current.

Unlike the out-of-plane spin current, the photo-driven *x*-component and *y*-component pseudo-spin currents are dominated by Rashba SOI. Therefore, the $S_x$ component photoconductivity $\eta_{abc}^{S_x}$ and $S_y$ component photoconductivity $\eta_{abc}^{S_y}$ should be larger than the $S_z$ component photoconductivity $\eta_{abc}^{S_z}$, which is demonstrated by Figures 3c and 3d. The maximum values of the $\eta_{abc}^{S_x}$ and $\eta_{abc}^{S_y}$ are nearly twice as that of $\eta_{abc}^{S_z}$.

Furthermore, the PSC direction can be effectively tuned by the polarization of incident light. For example, we have calculated the PSC for different polarization directions (in the *yz* plane) of the incident light at photon frequency $\hbar\omega = 0.9$ eV. Figures 3e and 3f show the Y-direction and Z-direction PSCs, respectively. Specifically, the spin polarization direction can be tuned in the $xz$ plane if measuring the Y-direction current, since the $S_y$ spin-polarized current is zero for $yz$ plane-polarized light. Compared with the Y-direction spin current, the Z-direction (ferroelectric polarization direction) spin current is one order of magnitude smaller, as indicated in Figure 3f.

***Ferroelectricity-driven switchable PSC.*** The above first-principles calculations and theoretical analyses illustrate the microscopic mechanism governing the nonlinear photo-driven PSCs. Because they are intimately related to the spin-texture and SOI, PSCs are inherently coupled with the intrinsic polarization of bulk ferroelectrics. This gives rise to ferroelectricity-induced switchable photo-driven spin current. In the following, we inspect the coupling between the ferroelectric order ($P_z$) and photo-driven PSC response.

The direction of spin-texture and the sign of photo-driven PSC susceptibility tensor will flip upon a ferroelectric polarization switch ($P_s$ to $-P_s$) in Figures 4a and 4c, which is because both the sign of Rashba SOI strength $\lambda_R$ and cubic Dresselhaus SOI strength $\lambda_{cd}$ will flip. Specifically, in Figure 4b, the spin texture of the conduction band is "clockwise" for ferroelectric with $P_s$. However, the spin texture of the same band is changed to "counterclockwise" after the polarization of ferroelectric flipping to $-P_s$, which is plotted in Figure 4d. Thus, under the same linearly polarized light, the direction of the photo-driven spin current will be rotated by 180° upon a ferroelectric polarization switch, as illustrated in Figures 4a and 4b. As shown in Figure 4e, the first-principles calculation confirms that the Y-axis photo-driven PSC can be switched by the ferroelectric polarization. The ferroelectric polarization in Figure 4e is normalized to the spontaneous polarization ($P_s$) of bulk GeTe. It is clear that the sign of photo-driven PSC susceptibility tensors $\eta_{xxY}^{S_z}$ and $\eta_{xxY}^{S_x}$ flips upon the ferroelectric polarization switch.

In summary, we have presented a giant, tunable photo-driven PSC in ferroelectrics with Rashba and cubic Dresselhaus SOI. We quantitatively show the generation of nonlinear DC spin current under linearly polarized light through the quantum perturbation theory. Using a prototypical ferroelectric material, bulk α-GeTe, we predict that PSC can be two orders of magnitude larger than the charge current. We have revealed the mechanism of such a giant PSC from the sizeable Rashba and cubic Dresselhaus SOIs. Finally, the PSC can be tuned and switched by both the light polarization and nonvolatile ferroelectric polarization, enriching the degrees of freedom to manipulate the spin current.

**Computational Details.**

***First-principles atomistic and electronic structure calculations.*** The First-principles calculations such as the ground state, crystal structure, and optical oscillate strength are mainly performed with the Vienna Ab-initio Simulation Package (VASP) [46] using the projector augmented wave [47] methods with the valence electrons of Ge and Te is described by the configurations ($4s^24p^2$) and ($5s^25p^4$), respectively. We use the plane-wave basis with an energy cutoff of 500 eV. The Perdew-Burke-Ernzerho (PBE) exchange-correlation functional [48] was used with spin-orbital coupling. The Structure optimizations were performed with a force criterion of 0.01 eV/Å. The Monkhorst-Pack k-point meshes of $8 \times 8 \times 8$ were adopted for the structural calculations of Bulk GeTe.

***First-principles nonlinear response calculations.*** To evaluate NLO responses, we calculate the derived formula based on the output such as velocity matrices, Berry curvature, etc from the first-principles DFT packages (e.g., VASP). The Monkhorst-Pack k-point meshes of $64 \times 64 \times 64$ because of the giant SOI (see the convergence test in the supporting information).

FIGURES:

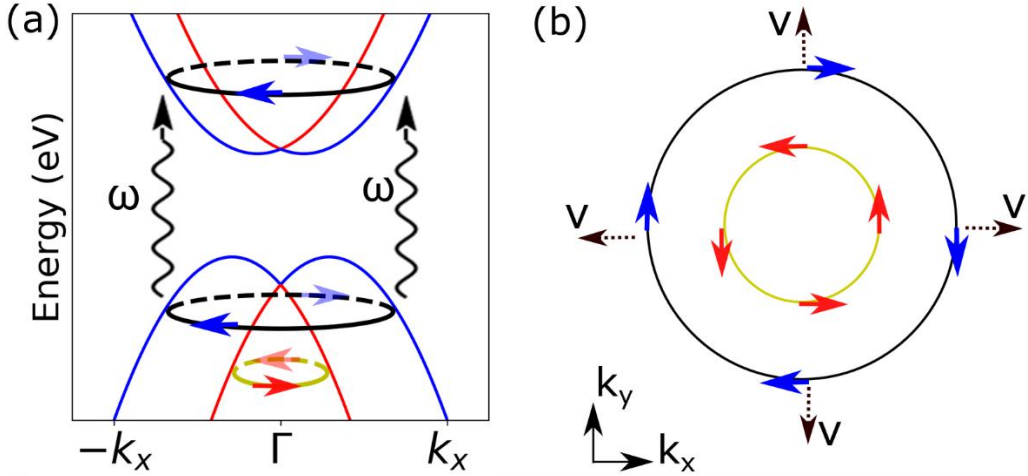

**Figure 1. Diagrams for the mechanism of photo-driven spin current.** (a) The diagram of Rashba bands with spin-texture. The blue and red arrows indicate the spin directions. Together with the arrows, the blue and red bands characterize the spin rotation directions as "clockwise" and "counterclockwise", respectively. (b) The group velocity of excited carriers with different spins. The excited electrons in conduction bands with different spins travel in opposite directions, forming a net spin-polarized current.

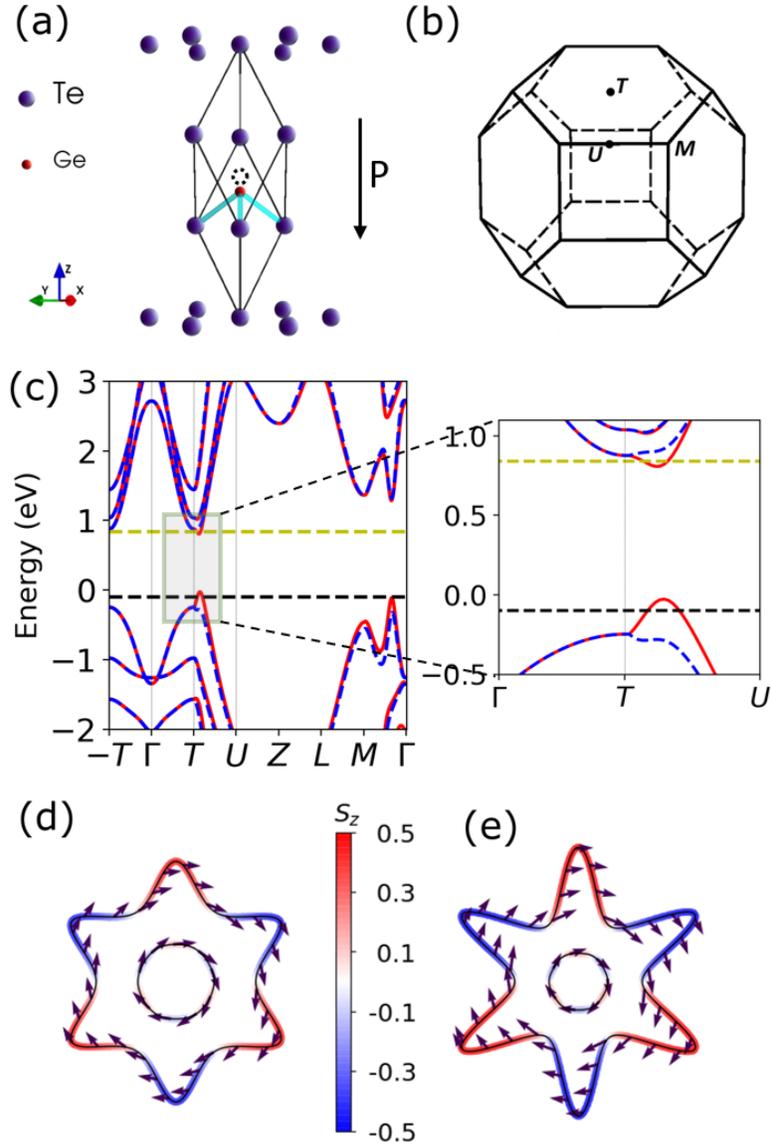

**Figure 2. Rashba and cubic Dresselhaus SOI in bulk α-GeTe.** The atomic structure (a) and the first Brillouin zone (b) of bulk GeTe. The dashed circle in (a) represents the Ge composition in its inversion symmetric structure. (c) The bandstructure along the main symmetry lines in the first Brillouin zone. The spin texture of valence bands (d) at the energy marked by the black dashed line in (c), and the spin texture of the conduction band (e) at the energy marked by the yellow dashed line in (c). The color bar in (d) and (e) indicates the normalized z-direction component of spin $S_z$.

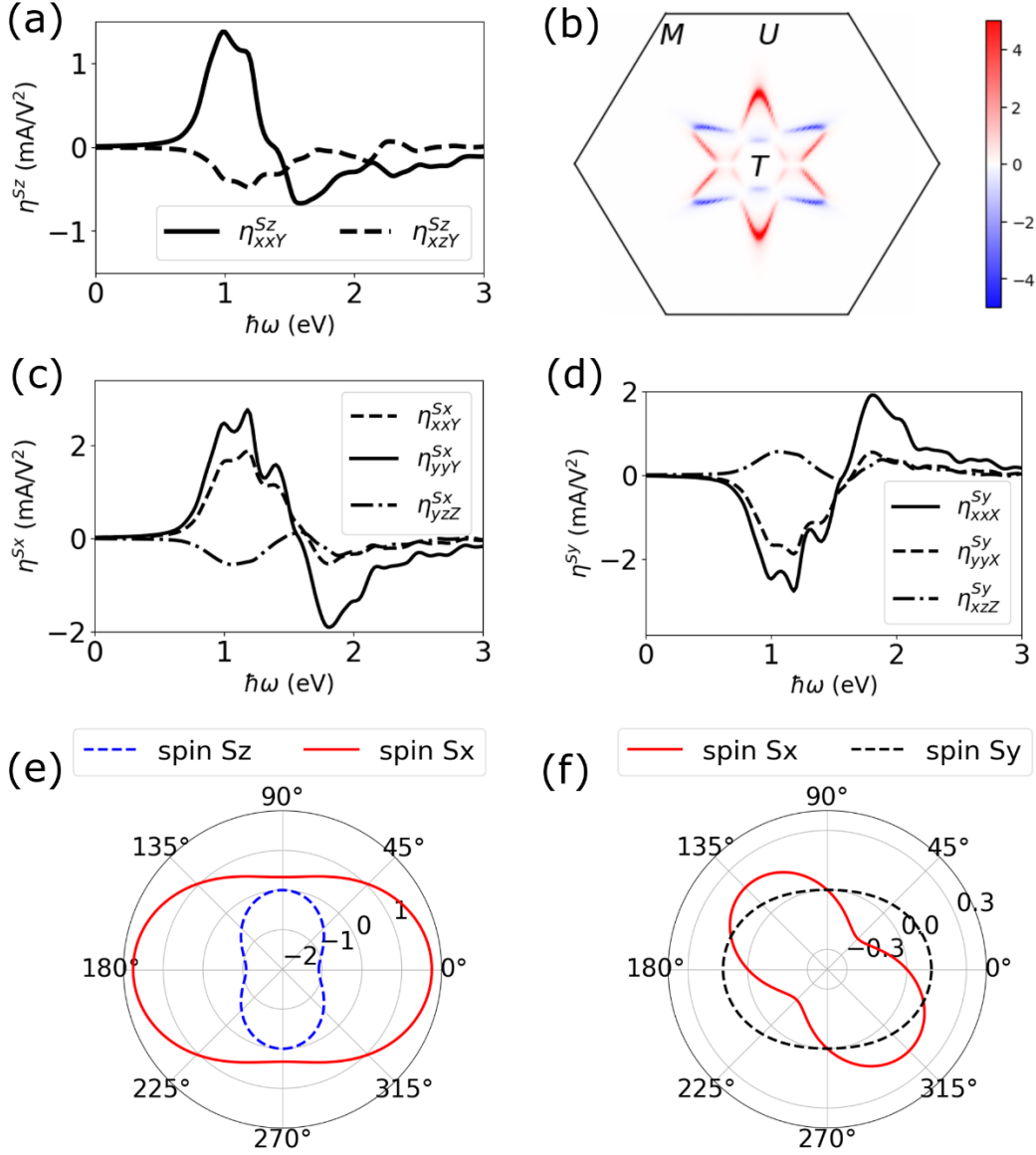

**Figure 3. Giant photo-driven spin current in bulk GeTe**. (a) Photo-driven spin current conductivity for the $z$-component spin ($S_z$). (b) Contour plot of calculated $S_z$ component photoconductivity $\eta^{S_z}_{xxY}$ with SOI on the (111) plane of the reciprocal space. The photo-driven spin current conductivity of the $x$-component spin (c) and $y$-component spin (d). The $Y$-direction (e) and $Z$-direction (f) components of the transported spin current for different polarization of light in the (100) plane; The 0-degree represents the $x$-direction polarized light and the incident photons are at the energy of 0.9 eV.

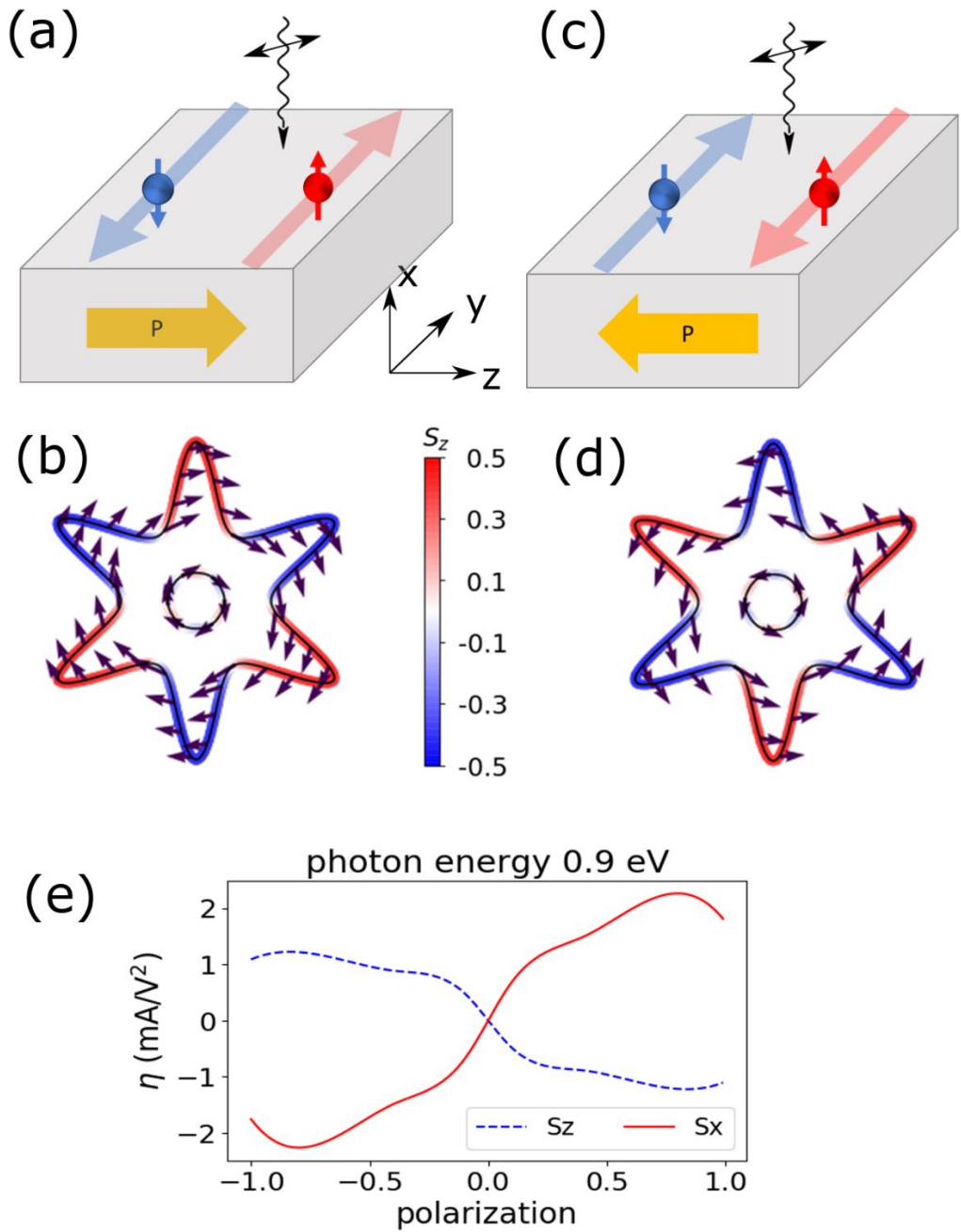

**Figure 4. Switchable photo-driven pure spin current**. (a) (c), The schematic plots for the spin current of opposite ferroelectric polarization under the same polarized light. (b) and (d) are the spin texture of the conduction band for positive polarization (a) and negative polarization (c) respectively. (d) First-principle calculated spin current photoconductivity $\eta^{S_z}_{xxY}$ (blue dash line) and photoconductivity $\eta^{S_x}_{xxY}$ (red solid line) with different ferroelectric polarization.

**Supporting Information**.

Supporting Information include. The Supporting Information is available free of charge on the xx.

The demonstration of type II spin textures of conduction and valence bands and the First-principles calculation convergence test. (PDF)

**Corresponding Author**

*(R.F.) E-mail: ruixiangfei@gmail.com
*(Y.L.) E-mail: lyang@physics.wustl.edu

**Notes**

The authors declare no competing financial interest.

**ACKNOWLEDGMENT**

R.F., L.Z., and L.Y. are supported by the Air Force Office of Scientific Research (AFOSR) grant No. FA9550-20-1-0255 and the National Science Foundation (NSF) CAREER grant No. DMR-1455346. The computational resources are provided by the Stampede of Teragrid at the Texas Advanced Computing Center (TACC) through XSEDE.

TOC figure

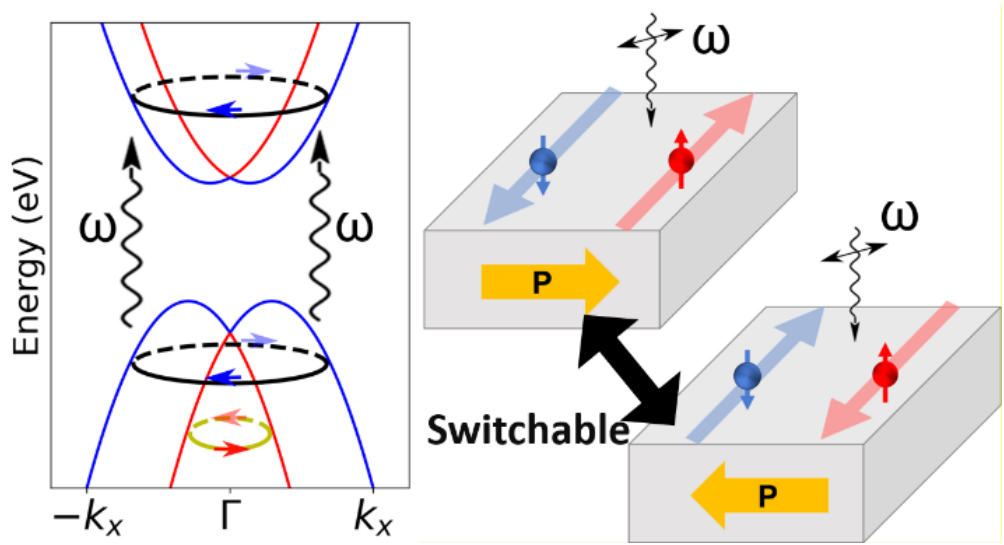